\documentclass[graybox]{svmult}

\usepackage{type1cm}        
\usepackage{makeidx}         
\usepackage{graphicx}        
\usepackage{multicol}        
\usepackage[bottom]{footmisc}

\usepackage{newtxtext}       %
\usepackage{newtxmath}       
\usepackage{pdfx}
\usepackage[numbers,sort&compress]{natbib}
\usepackage{balance}

\usepackage{bbold}

\newcommand{\myparatight}[1]{\smallskip\noindent{\bf {#1}:}~}

\makeindex 

\begin{document}
%
\title*{10 Security and Privacy Problems in Large Foundation Models}

\author{Jinyuan Jia, Hongbin Liu, Neil Zhenqiang Gong}
\institute{Jinyuan Jia \at Duke University, Durham, NC 27708, \email{jinyuan.jia@duke.edu}
\and Hongbin Liu \at Duke University, Durham, NC 27708, \email{hongbin.liu@duke.edu}
\and Neil Zhenqiang Gong \at Duke University, Durham, NC 27708, \email{neil.gong@duke.edu}}

\maketitle

 \abstract*{Foundation models--such as GPT, CLIP, and DINO--have achieved revolutionary progress in the past several years and are commonly believed to be a promising approach for general-purpose AI. In particular, self-supervised learning is adopted to pre-train a foundation model using a large amount of unlabeled data. A pre-trained foundation model is like an ``operating system'' of the AI ecosystem. Specifically, a foundation model can be used as a feature extractor for many downstream tasks with little or no labeled training data.  Existing studies on foundation models mainly focused on pre-training  a better foundation model to improve its performance on downstream tasks in non-adversarial settings, leaving its security and privacy in adversarial settings largely unexplored. A security or privacy issue of a pre-trained foundation model leads to a single point of failure for the AI ecosystem. In this book chapter, we discuss 10 basic security and privacy problems for the pre-trained foundation models, including six confidentiality problems, three integrity problems, and one availability problem. For each problem, we discuss potential opportunities and challenges. We hope our book chapter will inspire future research on the security and privacy of foundation models. }

 \abstract{Foundation models--such as GPT, CLIP, and DINO--have achieved revolutionary progress in the past several years and are commonly believed to be a promising approach for general-purpose AI. In particular, self-supervised learning is adopted to pre-train a foundation model using a large amount of unlabeled data. A pre-trained foundation model is like an ``operating system'' of the AI ecosystem. Specifically, a foundation model can be used as a feature extractor for many downstream tasks with little or no labeled training data.  Existing studies on foundation models mainly focused on pre-training  a better foundation model to improve its performance on downstream tasks in non-adversarial settings, leaving its security and privacy in adversarial settings largely unexplored. A security or privacy issue of a pre-trained foundation model leads to a single point of failure for the AI ecosystem. In this book chapter, we discuss 10 basic security and privacy problems for the pre-trained foundation models, including six confidentiality problems, three integrity problems, and one availability problem. For each problem, we discuss potential opportunities and challenges. We hope our book chapter will inspire future research on the security and privacy of foundation models. }

\section{Introduction}
Deep neural network based supervised learning has achieved tremendous success in the last decade. However, its major defect is that it relies on a large amount of labeled data for each machine learning task. As an alternative, self-supervised learning~\cite{hadsell2006dimensionality,devlin2019bert,chen2020simple,radford2021learning,bommasani2021opportunities} has attracted more and more attention and is believed to be a promising AI paradigm that can address the defect in supervised learning. In particular, self-supervised learning aims to pre-train a foundation model (also called \emph{encoder}\footnote{We use foundation model and encoder interchangeably in this chapter.}) on a large amount of unlabeled data that can be used as a feature extractor for many downstream tasks. For instance, in computer vision domain, self-supervised learning can pre-train an image encoder on unlabeled images, or both an image and a text encoder on  unlabeled (image, text) pairs. The pre-trained image encoder (and text encoder) can be used to build a downstream classifier for a downstream task that has little or even no labeled training examples. In natural language processing domain, self-supervised learning aims to pre-train a language model on unlabeled texts that can be further fine-tuned for a natural language downstream task. In graph domain, self-supervised learning can pre-train a graph encoder on unlabeled graphs that can be used for many graph downstream tasks such as node classification and graph classification. 

Many recent studies showed that pre-trained encoders can achieve state-of-the-art results on many downstream tasks in various domains. For instance, researchers from OpenAI collected 400 million (image, text) pairs from the Internet and used them to pre-train both an image and a text encoder~\cite{radford2021learning}. They found that classifiers built upon those encoders with no labeled training data can achieve classification accuracies that are comparable to classifiers learnt by supervised learning with a large amount of labeled training data. As encoders are pre-trained on a large amount of unlabeled data, the required resources (e.g., computation, memory, storage resources) are usually unaffordable for individuals. Therefore, a typical paradigm is that a resourceful tech or AI research company such as Google, Facebook, and OpenAI pre-trains encoders and then shares them with its customers. Many self-supervised learning algorithms have been proposed to pre-train encoders such that the accuracies of downstream classifiers built upon the pre-trained encoders can be improved. However, security and privacy of self-supervised learning receive little attention from both academia and industry. 

In this book chapter, we discuss 10 security and privacy problems faced by foundation models pre-trained by self-supervised learning. Roughly speaking, information security and privacy concerns about \emph{confidentiality}, \emph{integrity}, and \emph{availability}. Specifically, confidentiality means that information is protected from unauthorized access. In foundation models, both pre-training dataset and pre-trained encoders could be proprietary. Along those two dimensions, we discuss six confidentiality problems,  four of which are related to the pre-training dataset and two of which are related to the pre-trained encoders. In particular, the pre-training dataset could be collected from the Internet. Thus, an individual may be interested in whether his/her data is collected and used to pre-train encoders without his/her authorization. The first confidentiality problem discussed in this book chapter is \emph{data tracing/auditing} which could be used by an individual to trace/audit the usage of his/her data in self-supervised learning. The pre-training dataset could also be private data such as healthcare data. An attacker could perform various attacks to compromise the confidentiality of the private pre-training dataset. In particular, the attacker could perform: 1) \emph{membership inference attacks} to infer whether a data sample is in the pre-training dataset of an encoder, 2) \emph{reconstruction attacks} to reconstruct  typical data samples in the pre-training dataset, and 3) \emph{attribute/property inference attacks} to infer sensitive attributes of a sample  or  the pre-training dataset. These attacks pose severe security and privacy concerns to foundation models. Therefore, an interesting future research is to study those attacks and their defenses.

The pre-trained encoders represent the intellectual property of an encoder provider. When a pre-trained encoder is provided as a service to customers via an API, an attacker may try to steal hyperparameters (e.g., encoder architecture) of the encoder. We call such  attacks  \emph{encoder hyperparameter stealing attacks}. Instead of stealing hyperparameters, an attacker can also try to steal parameters of the encoder. We call them \emph{encoder parameter stealing attacks}. These two attacks are largely unexplored for self-supervised learning. 

We also discuss three integrity problems for foundation models. Integrity means that information is protected from unauthorized alternation. A typical machine learning system involves two phases, namely training phase and testing phase. Therefore, an attacker can compromise the integrity of self-supervised learning in  1) training phase, 2) testing phase, and 3) both training and testing phases. We call the attacks that compromise the training phase \emph{poisoning attacks}, call the attacks that compromise the testing phase \emph{evasion attacks}, and call the attacks that compromise both training and testing phases \emph{backdoor attacks}. To compromise the training phase, an attacker can poison the pre-training dataset (a.k.a. \emph{data poisoning attacks}) or tamper the training process (a.k.a. \emph{model poisoning attacks}). To compromise the testing phase, an attack can tamper the testing data.  These three attacks and their defenses for self-supervised learning are largely unexplored, which we believe is an interesting future research direction. 

Finally, we discuss one availability problem for foundation models. Availability means that a system can be accessed by an intended user timely. In particular, we discuss resource depletion attacks, in which an attacker tries to reduce the convergence rate and increase the computation time of pre-training an encoder via poisoning the pre-training dataset or  increase the inference  time of a deployed encoder via carefully crafting testing inputs.

\noindent
{\bf Roadmap:} In Section~\ref{relatedwork}, we discuss background knowledge on foundation models. We respectively discuss six confidentiality problems, three integrity problems, and one availability problem of foundation models in Section~\ref{confidentiality},~\ref{integrity}, and~\ref{availability}. Finally, we conclude in Section~\ref{conclusion}.

\section{Background on Self-supervised Learning}
\label{relatedwork}
We discuss self-supervised learning in natural language processing (NLP), computer vision (CV), and graph domains. Roughly speaking, self-supervised learning aims to use a large amount of unlabeled data to pre-train an encoder. For instance, the unlabeled data could be unlabeled texts collected from Wikipedia in NLP domain,  unlabeled images or (image, text) pairs in CV domain, and unlabeled graphs in graph domain. Suppose we have a downstream task that has a small amount of or no labeled data.  We can use the encoder as a feature extractor to build a downstream classifier for the downstream task. In practice, the encoder could be pre-trained by resourceful tech companies (e.g., Google, Facebook, and Amazon) or AI research company (e.g., OpenAI) and then is shared with downstream customers.

\subsection{Self-supervised Learning in NLP}

\subsubsection{Pre-training a Language Model}
In NLP domain, self-supervised learning aims to pre-train a language model using a large amount of unlabeled texts~\cite{devlin2019bert}. Then, we can fine-tune the pre-trained language model for a downstream task. Essentially, the language model is trained to learn the distribution of word sequence (e.g., phrases and sentences) using the collected corpus, e.g., BookCorpus or English Wikipedia. We take BERT~\cite{devlin2019bert}, a state-of-the-art language model, as an example to dig into the details.

The input of BERT could be either a single sentence or a pair of sentences, where a sentence can be an arbitrary span of contiguous text. Given the input, BERT further converts it to a sequence of tokens and outputs an embedding vector for each token. To train the language model, BERT considers two tasks. In the first task, we randomly mask a certain fraction of input tokens and then train BERT to predict those missing tokens. The second task is next sentence prediction. Specifically, given a pair of sentences, BERT is trained to predict whether the second sentence is the next sentence of the first sentence. 

\subsubsection{Fine-tuning a Language Model for a Downstream Task}
The pre-trained language model can be used for many natural language downstream tasks such as question answering and sentiment analysis. For instance, to fine-tune BERT for a specific downstream task, we can first add the task dependent inputs and outputs to BERT and then fine-tune all the parameters in an end-to-end manner.

\subsection{Self-supervised Learning in CV}
We first introduce how to pre-train an image encoder using unlabeled images or (image, text) pairs. Then, we discuss how to leverage the image encoder as a feature extractor for many downstream tasks.

\subsubsection{Pre-training an Image Encoder (and a Text Encoder)}

\myparatight{Unlabeled images} Many self-supervised learning algorithms~\cite{hadsell2006dimensionality,grill2020bootstrap,he2020momentum,chen2020simple} have been proposed to use unlabeled images to pre-train an image encoder. Among them,  contrastive learning~\cite{he2020momentum,chen2020simple} achieves state-of-the-art performance. 
Roughly speaking, the idea of contrastive learning is to learn feature representations that are invariant to image augmentations. To reach the goal, contrastive learning updates parameters of an image encoder to minimize the contrastive loss. In particular, the contrastive loss is small if the image encoder outputs similar (or dissimilar) feature representations for two augmented images generated from the same image (or different images). To dig into the details, we take SimCLR~\cite{chen2020simple} as an example to illustrate contrastive learning for its simplicity and effectiveness. 

The framework of SimCLR contains three modules: 1) a stochastic data augmentation, 2) an image encoder, and 3) a projection head. The stochastic data augmentation module is used to generate a randomly augmented image for an image via applying a composition of several data augmentations, e.g., random cropping and random color distortions. The image encoder produces a feature representation for an (augmented) image. The projection head projects the feature representation to the space where the contrastive loss is defined.
In each training step, SimCLR randomly samples $N$ unlabeled images to update the parameters of the image encoder and the projection head. In particular, SimCLR applies the stochastic data augmentation module to each image twice to generate $2 \cdot N$ augmented images. For simplicity, we use $\mathbf{x}_1, \mathbf{x}_2, \cdots, \mathbf{x}_{2\cdot N}$ to denote those augmented images where $\mathbf{x}_{2\cdot i -1}$ and $\mathbf{x}_{2\cdot i}$ ($i=1,2,\cdots, N$) form a positive pair. Suppose we respectively use $f$ and $g$ to denote the image encoder and projection head. Moreover, we define $\ell(i,j)$ as follows:
\begin{align}
    \ell(i,j) = - \log(\frac{\exp(sim(g\circ f(\mathbf{x}_i), g\circ f(\mathbf{x}_j))/\tau)}{\sum_{k=1}^{2\cdot N} \mathbb{I}(k \neq i) \cdot \exp(sim(g\circ f(\mathbf{x}_i), g\circ f(\mathbf{x}_k))/\tau) } ),
\end{align}
where $\mathbb{I}$ is an indicator function, $\tau$ is a temperature parameter, and $g \circ f$ represents the composition of projection head $g$ and the image encoder $f$. In each training step, SimCLR updates the parameters of the image encoder $f$ and the projection head $g$ to minimize the average contrastive loss of the $2 \cdot N$ augmented images:
\begin{align}
    \min_{g,f}\mathcal{L} = \frac{1}{2\cdot N} \sum_{i=1}^{N}(\ell (2\cdot i-1, 2 \cdot i) + \ell (2\cdot i, 2 \cdot i - 1)).
\end{align}

\myparatight{Unlabeled (image, text) pairs} Self-supervised learning could also pre-train an image encoder~\cite{li2017learning,radford2021learning} via leveraging supervisory signals in the text. In particular, we can first collect some images along with texts (i.e., (image, text) pairs) from the Internet as the pre-training dataset, and then use the collected pre-training dataset to jointly pre-train an image encoder and a text encoder. Recently, Radford et al.~\cite{radford2021learning} proposed Contrastive Language–Image Pre-training (CLIP) which uses 400 million (image, text) pairs collected from the Internet as the pre-training dataset to pre-train an image encoder and a text encoder. Radford et al. showed that the pre-trained encoders can achieve state-of-the-art performance on many downstream tasks. Next, we briefly show the general idea behind CLIP. 

Suppose we have $N$ (image, text) pairs which are denoted as $(\mathbf{x}_1, T_1),$ $ (\mathbf{x}_2, T_2), \cdots ,$ $ (\mathbf{x}_N, T_N)$. CLIP tries to jointly optimize the image encoder and the text encoder via predicting which image pairs with which text. In particular, for each image $\mathbf{x}_i$, CLIP aims to make the output of the image encoder on $\mathbf{x}_i$ similar (or dissimilar) to the output of the text encoder on $T_i$ (or $T_j, j \neq i$). The similarity between the output of the image encoder and that of the text encoder is measured by the cosine similarity. Moreover, CLIP uses random square crop to augment images in the training process.

\subsubsection{Applying an Image Encoder (and a Text Encoder) to Downstream Tasks}

We consider two scenarios based on whether a downstream task (we consider classification task for simplicity) has labeled images. In particular, we can use an image encoder as a feature extractor to fine-tune a (or assemble a zero-shot) downstream classifier for a downstream task if it has (or does not have) labeled images. Next, we respectively introduce how to build and use a downstream classifier in those two scenarios. 
    
    \myparatight{Fine-tuning a downstream classifier} Suppose we have a set of labeled images (called \emph{downstream dataset}) in a downstream task. To build a downstream classifier, we can add one or several linear layers (with certain activation functions) on top of an image encoder. Note that the softmax activation function is usually applied to the last linear layer. Then, we can freeze the parameters of the image encoder and fine-tune the downstream classifier on the downstream dataset. The fine-tuned downstream classifier can be used to predict a label for a testing image. 
    
    \myparatight{Assembling a zero-shot downstream classifier} Suppose we have a set of $c$ desired class names (e.g., \{``dog'', ``cat'', $\cdots$, ``deer'' \}) of a downstream task, but do not have any labeled images. Moreover, we also have both an image encoder and a text encoder. We can assemble a zero-shot classifier in this scenario. A naive solution is that we directly use the class names to assemble a zero-shot classifier. However, the zero-shot classifier assembled in this way may achieve sub-optimal classification accuracy. The reason is that the texts in the pre-training dataset are usually sentences, which means a single word class name may be rarely paired with an image in the pre-training dataset. Therefore, we can add a descriptive context to each class name. For example, Radford et al.~\cite{radford2021learning} found that ``A photo of a \{class name\}'' is a good template to be used for CLIP's encoders. To assemble a zero-shot classifier, we first create $c$ context sentences via adding a descriptive context to each class name. Then, we use the text encoder to compute a feature representation for each context sentence. Similarly, given a testing image, we use the image encoder to compute a feature representation for it. Then, the testing image is predicted to belong to the class whose context sentence's feature representation has the largest similarity with the testing image's feature representation.

\subsection{Self-supervised Learning in Graph}
Self-supervised learning has also been extended to graph domain~\cite{qiu2020gcc,you2020graph,zhu2021graph}. In particular, the goal is to pre-train a graph encoder using a large amount of unlabeled graphs. Similar to NLP and CV domains, the pre-trained graph encoder can be used as a feature extractor for many downstream tasks in graph learning.

\subsubsection{Pre-training a Graph Encoder}
Multiple studies~\cite{qiu2020gcc,you2020graph,zhu2021graph} generalized contrastive learning from CV domain to graph domain. In particular, the goal is to pre-train an image encoder such that it outputs similar (or dissimilar) feature representations for two subgraphs that are generated from the same (or different) graph(s). Next, we discuss GraphCL~\cite{you2020graph} which generalizes SimCLR~\cite{chen2020simple} from CV domain to graph domain. 

Similar to SimCLR, GraphCL also has three major modules: a graph data augmentation module, a GNN-based graph encoder, and a projection head. The graph data augmentation module produces a randomly augmented graph from a given graph via applying a sequence of graph manipulation operations, e.g., randomly discard a certain fraction of nodes as well as their connections, randomly add or delete some edges, sample a subgraph via random walk. The GNN-based graph encoder outputs a feature representation for a graph or an augmented graph. The projection head maps the feature representation to the space where the contrastive loss is defined. Given a mini-batch of $N$ graphs, GraphCL generates $2 \cdot N$ augmented graph via applying the graph data augmentation module to each graph in the mini-batch twice. For simplicity, we use $G_{i,1}, G_{i,2}$ to denote the two augmented graphs generated from the $i$th graph in the mini-batch. Note that $G_{i,1}$ and $ G_{i,2}$ form a positive pair. Suppose we use $f$ and $g$ to respectively represent the GNN-based graph encoder and the projection head. Moreover, we define $\ell(i)$ as follows:
\begin{align}
    \ell(i) = - \log(\frac{\exp(sim(g\circ f(G_{i,1}), g\circ f(G_{i,2}))/\tau)}{\sum_{k=1}^{ N} \mathbb{I}(k \neq i) \cdot \exp(sim(g\circ f(G_{i,1}), g\circ f(G_{k,2}))/\tau) } ),
\end{align}
where $\mathbb{I}$ is an indicator function, $\tau$ is a temperature parameter, and $g \circ f$ represents the composition of $g$ and $f$. GraphCL tries to minimize $\frac{1}{N}\sum_{i=1}^N \ell(i)$ to train the GNN-based graph encoder $f$ and the projection head $g$ in each training step.

\subsubsection{Applying a Graph Encoder to Downstream Tasks}
The pre-trained graph encoder can be used as a feature extractor for many downstream tasks such as node classification and graph classification, which are two basic graph learning tasks. 

\myparatight{Graph classification} Suppose we have a small amount of labeled graphs. We can use the graph encoder to extract a feature representation for each of them. Then, we can use the standard supervised learning to train a classifier using those feature representations as well as the corresponding labels. Given a testing graph, we can first use the graph encoder to extract a feature representation and then use the classifier to predict a label for the feature representation. The predicted label is viewed as the prediction result for the testing graph.

\myparatight{Node classification} Suppose we have a set of labeled nodes in a graph, we wish to classify the remaining unlabeled nodes. For each node, we can extract a subgraph centered at the node. Given the extracted subgraphs of the labeled nodes, we can use a graph encoder to compute a feature representation for each of them. Then, we can train a classifier using those feature representations as well as the corresponding labels by standard supervised learning. Given an unlabeled node, we can first extract its subgraph, then use the graph encoder to compute a feature representation, and finally use the classifier to make a prediction for the feature representation.

\section{Six Problems on Confidentiality}
\label{confidentiality}
In foundation models, the encoder provider may wish to keep the pre-training dataset and/or encoder confidential since they may represent the model provider's intellectual property, while individuals may desire protection and authorized access of their data in the pre-training dataset. We discuss six potential confidentiality problems,  four of which are related to the pre-training dataset and two of which are related to the encoder. 

\myparatight{Data Tracing/Auditing} Individuals have the right to know how their data is used as per the requirements of data protection laws and regulations such as GDPR~\cite{EU_GDPR}. Unauthorized use of individuals' data may violate copyright regulations and/or increase privacy risks of individuals. For instance, Ever was asked by FTC to delete models trained on unauthorized individual data~\cite{FTC_settlement}. Therefore, it is essential for an individual to be able to trace/audit the use of his/her data in machine learning systems. 
In particular, an individual should be able to verify whether his/her data was used to train a machine learning model. 
When an individual requests a model provider to delete his/her data from a model, data tracing/auditing can also be used to verify whether the model provider follows the request. Some studies~\cite{sablayrolles2020radioactive,sommer2020towards} developed data tracing/auditing techniques for machine learning classifiers. For instance, Sablayrolles et al.~\cite{sablayrolles2020radioactive} proposed to add a radioactive mark to training images (called radioactive images) to detect whether those radioactive images are used to train a machine learning classifier. In particular, by leveraging statistical analysis, they can detect the use of radioactive images in training a machine learning classifier with high confidence. Data tracing/auditing has also been studied for text-generation model~\cite{song2019auditing} in which a user could audit whether his/her words are used to train a publicly available text-generation model without his/her authorization.

Self-supervised learning usually uses a large amount of public data collected from the Internet to pre-train an encoder. For instance, Radford et al.~\cite{radford2021learning} collected 400 million (image, text) pairs from the Internet and used them to pre-train CLIP's encoders. An individual could use data tracing/auditing techniques to verify whether his/her data has been used to pre-train an encoder without his/her authorization. For instance, Liu et al.~\cite{liu2021encodermi} proposed a method for data tracing in contrastive learning, which can be used by an individual to verify whether its public data on the Internet was used to pre-train encoders (e.g., CLIP). However, this method does not have formal guarantees, e.g., there is no formal statistical confidence guarantee for the data tracing decisions. 
Moreover, an individual could also proactively perturb his/her data to facilitate data tracing/auditing. For instance, an individual could add carefully crafted marks to his/her images before uploading them to the Internet.
We believe it is an interesting future direction to develop data tracing/auditing techniques with formal guarantees for self-supervised learning.

\myparatight{Membership Inference Attacks} In membership inference attacks, an attacker aims to infer whether a data sample is used to train a target machine learning model. In particular, the data sample is called a \emph{member} (or \emph{non-member}) of the target machine learning model if it is used to (or not used to) train the target machine learning model. Membership inference can be used for data tracing/auditing, though without formal guarantees. 
Membership inference poses severe security and privacy concerns for machine learning models when a machine learning model is trained on proprietary, sensitive data such as healthcare data. Moreover, membership inference attack can be used as a step stone for other attacks such as attribute inference attacks.  

In the past several years, membership inference attacks  have been studied for image classifiers~\cite{shokri2017membership,yeom2018privacy, salem2018ml,nasr2019comprehensive, song2019privacy, choquette2021label, li2020label,hui2021practical}, image generative models~\cite{hayes2019logan, chen2020gan}, language models~\cite{carlini2020extracting}, and graph neural networks~\cite{he2021stealing}. For instance, Shokri et al.~\cite{shokri2017membership} extended membership inference attack~\cite{homer2008resolving,backes2016membership} to machine learning classifier, where an attacker trains a binary attack classifier to infer whether a data sample is in the training dataset of a target classifier by leveraging the confidence scores predicted by the target classifier for the data sample.  
Carlini et al.~\cite{carlini2020extracting} showed that an attacker can infer whether a text is used to train a language model, e.g., a text is likely to be a member of a language model's training data if the language model assigns high likelihood for the text. He et al.~\cite{he2021stealing} showed that an attacker can infer whether there exists an edge between two nodes in a graph via black-box access to a graph neural network trained on the graph.  

Many methods~\cite{shokri2017membership,salem2018ml,nasr2018machine,jia2019memguard} have been proposed to defend against membership inference attacks such as regularization~\cite{shokri2017membership,salem2018ml}, ensemble  method~\cite{salem2018ml}, differential privacy~\cite{abadi2016deep}, game-theoretic method~\cite{nasr2018machine}, adversarial example based method~\cite{jia2019memguard}, and so on~\cite{song2021systematic}. For instance, 
Jia et al.~\cite{jia2019memguard} proposed MemGuard, which leverages adversarial examples to defend against membership inference attacks. In particular, the idea is to add carefully crafted noise to the confidence scores predicted by a target classifier for a data sample such that an attacker's classifier cannot correctly predict whether the data sample is a member of the target classifier's training dataset. 

Recently, two concurrent works~\cite{liu2021encodermi, he2021quantifying} studied membership inference attacks to contrastive learning. In particular, Liu et al.~\cite{liu2021encodermi} developed the first membership inference attack, namely EncoderMI, to image encoders pre-trained by contrastive learning. They showed that an attacker can exploit the overfitting of an image encoder on a pre-training dataset to infer whether an unlabeled image is in the pre-training dataset. The idea is that an overfitted image encoder is more likely to produce similar (or dissimilar) feature representations for two augmented images that are generated from the same (or different) image(s). He \& Zhang et al.~\cite{he2021quantifying} explored the effectiveness of existing membership inference attacks for contrastive learning. In particular, they consider the scenario where the pre-training dataset is the same as the downstream dataset. In other words, they first use a labeled dataset to pre-train an image encoder and then use the labeled dataset and the image encoder as a feature extractor to train a downstream classifier. They found that the downstream classifier is less vulnerable to existing membership inference attacks. Liu et al.~\cite{liu2021encodermi} also found that existing defenses against membership inference attacks to classifiers are insufficient for pre-trained encoders. For instance, Liu et al.~\cite{liu2021encodermi} explored early stopping~\cite{song2021systematic} as a countermeasure and found that it achieves a tradeoff between the membership inference accuracy of EncoderMI and the utility of the image encoder. Therefore, an interesting future research direction is to explore how to defend against membership inference attacks to pre-trained image encoders.

\myparatight{Reconstruction Attacks} In reconstruction attacks, an attacker tries to reconstruct some typical training samples. Those attacks are also known as model inversion attacks and have been widely studied for classifiers~\cite{fredrikson2015model,wu2016methodology,hidano2018model,yeom2018privacy,yang2019adversarial,mehnaz2020black}. For instance,  Fredrikson et al.~\cite{fredrikson2015model} studied model inversion attacks to decision trees and neural networks. In particular, they found that, with access to an  individual's name and facial recognition model, an attacker could recover images of the individual's faces. Model inversion attacks have also been studied for language models~\cite{carlini2020extracting}. For instance, Carlini et al.~\cite{carlini2020extracting} showed that training examples of a pre-trained language model can be recovered via querying it. Their attacks are validated on GPT-2~\cite{radford2019language}, a language model trained on texts collected from the Internet by OpenAI.

To the best of our knowledge, reconstruction attacks have not been studied for self-supervised learning in image and graph domains. With either black-box or white-box access to an encoder, an attacker could try to perform reconstruction attacks to it. In particular, the attacker tries to recover some training inputs in the pre-training dataset. For instance, suppose an attacker can access both the text encoder and image encoder pre-trained on (image, text) pairs, the attacker could try to recover the image of an individual by his/her name. In particular, the attacker could try to craft an image whose feature representation produced by the image encoder has the largest similarity with the feature representation of ``A photo of \{individual's name\}" produced by the text encoder. When having access to the non-private part of an image, the attacker could try to reconstruct its private part; when having a blurred version of an image, an attacker could try to reconstruct the image; when having a partial graph, an attacker could try to reconstruct the remaining links in the graph.

\myparatight{Attribute/Property Inference Attacks} In attribute inference attacks, an attacker aims to infer sensitive attributes of a sample   via either black-box or white-box access to a  model and  the knowledge of other non-sensitive attributes of the sample. Attribute inference attacks were extensively studied in online social networks, which aim to infer users' private attributes (e.g., age, gender, location) using their publicly available data (e.g., page likes, rating scores) on social networks~\cite{gong2014joint,gong2016you,jia2017attriinfer,gong2018attribute}. Attribute inference attacks have also been extended to machine learning including regression~\cite{fredrikson2014privacy}, classification~\cite{fredrikson2015model,mehnaz2020black}, and federated learning~\cite{melis2019exploiting}. For instance, Fredrikson et al.~\cite{fredrikson2014privacy} developed attribute inference attacks to linear regression. In particular, they showed that an attacker could infer a patient's genetic markers when the attacker knows the patient's background and stable dosage. 
Instead of inferring sensitive attributes of an individual sample, property inference attacks~\cite{ateniese2015hacking,ganju2018property} aim to infer sensitive properties of the training dataset used to train a target model. When a target model is a commercial product that stays ahead of the competition, property inference attacks may allow competitors to uncover the reason via inferring the properties of the training dataset of the target model. Those attacks have been studied for various machine learning classifiers such as SVMs and neural networks. For instance, Ateniese et al.~\cite{ateniese2015hacking} proposed to train a meta-classifier to predict whether the training dataset of a SVM classifier has a certain property or not.  Ganju et al.~\cite{ganju2018property} developed property inference attacks to fully connected neural networks, where the key insight is that fully connected neural networks are invariant under the permutation of neurons when they are represented by matrices.

Song et al.~\cite{song2020overlearning} showed that representations learnt by a neural network can be used to infer sensitive attributes of samples. In particular, an attacker can train a classifier which takes a sample's representation outputted by a target model as input and predicts the sample's sensitive attributes.  
He \& Zhang~\cite{he2021quantifying} generalized the attacks to contrastive learning. In particular, they found that contrastive learning is more vulnerable to such attacks than supervised learning. 
Adversarial example based defense such as AttriGuard~\cite{jia2018attriguard} could be extended to mitigate such attribute inference attacks. The key idea is to add carefully crafted perturbations to feature representations. 
An interesting future work is to further study attribute/property inference attacks in self-supervised learning,  
which we expect to attract more attention in the near future.

\myparatight{Encoder Hyperparameter Stealing Attacks} In hyperparameter stealing attacks, an attacker tries to  infer the hyperparameters (e.g., neural network architecture, learning rate, batch size) that are used to train a model. Those attacks have been studied in supervised learning in the past several years~\cite{wang2018stealing,oh2018towards,hua2018reverse,yan2020cache}. For instance,  Wang \& Gong~\cite{wang2018stealing} proposed hyparparameter stealing attacks which are applicable to a wide range of machine learning algorithms such as neural network, logistic regression, and support vector machine. Oh et al.~\cite{oh2018towards} showed that an attacker could infer various attributes (e.g., architecture and optimization procedure) of a neural network classifier via querying it with a sequence of inputs.

It remains an open question how to steal encoder hyperparameters (e.g., encoder architecture, data augmentation modules) in self-supervised learning. We note that there are at least two motivations for an attacker to perform such attacks. First, the encoder hyperparameters are indispensable components in self-supervised learning. However, it is computationally expensive to search  good hyperparameters. Second, the stolen encoder hyperparameters could be further used to assist the stealing of the encoder model parameters, which we will discuss in the next part. We note that how to protect the hyperparameters used to pre-train an encoder is also an interesting future research. 

\myparatight{Encoder Parameter Stealing Attacks} In parameter stealing attacks, an attacker tries to steal the parameters of a machine learning model. Multiple studies~\cite{lowd2005adversarial,tramer2016stealing,papernot2017practical,jagielski2020high,hua2018reverse,orekondy2019knockoff,chandrasekaran2020exploring,carlini2020cryptanalytic} found that, with a black-box access to a target classifier, it is feasible for an attacker to steal the exact parameters or train an accurate surrogate classifier via querying the target classifier. For instance, Tram{\`e}r~\cite{tramer2016stealing} developed model parameter stealing attacks for various machine learning models such as decision trees, logistic regressions, and support vector machines. They found that an attacker can steal the exact coefficients of a linear classifier or the paths of a decision tree. Jagielski et al.~\cite{jagielski2020high} proposed functionally-equivalent attack which can steal the exact parameters of a two-layer neural network. Chandrasekaran et al.~\cite{chandrasekaran2020exploring} analyzed the connection between model parameter stealing attacks and active learning. In particular, they found that techniques developed for active learning can be used to implement model parameter stealing attacks. We note that model parameter stealing attacks have also been studied for language models~\cite{devlin2018bert}. For instance, Krishna et al.~\cite{krishna2019thieves} showed that, with black-box access, an attacker could extract a NLP model fine-tuned on the pre-trained BERT model when he/she does not have any training data.  Many defenses~\cite{orekondy2019prediction,kariyappa2020defending,jia2021entangled} have been proposed to mitigate parameter stealing attacks. For instance, Orekondy et al.~\cite{orekondy2019prediction} proposed to add carefully crafted noise to confidence scores outputted by a target classifier to attack the attacker's training objective. 

Parameter stealing attacks have not been studied for self-supervised learning. In other words, how to steal parameters of encoders pre-trained by self-supervised learning remains open questions. To answer the questions, interesting future research includes: 1) generalizing existing parameter stealing attacks or developing new parameter stealing attacks to steal parameters of encoders pre-trained by self-supervised learning, and 2) generalizing existing defenses or developing new defenses to protect parameters of encoders. The key challenge is how to exploit the unique characteristics of self-supervised learning when studying parameter stealing attacks as well as their defenses.

\section{Three Problems on Integrity}
\label{integrity}

There are two phases in machine learning: training phase and testing phase. Therefore, an attacker can compromise the integrity of machine learning in either one or both phases. More specifically, an attacker can compromise: 1) both training and testing phases (a.k.a. backdoor attacks), 2) training phase (a.k.a. poisoning attacks), and 3) testing phase (a.k.a. evasion attacks). An attacker can   compromise the training phase via tampering the training data or the training process, and can compromise the testing phase via tampering the testing data.

\myparatight{Backdoor Attacks} In backdoor attacks, an attacker compromises both the training and testing phases of a machine learning system such that the behaviors of the machine learning system is desired by the attacker. Backdoor attacks have been extensively studied for classifiers~\cite{liao2018backdoor,yao2019latent,saha2020hidden,turner2019label,li2019invisible,tan2019bypassing,liu2020reflection,salem2020dynamic,gu2017badnets,chen2017targeted,liutrojaning2018,dai2019backdoor,bagdasaryan2020backdoor,zhang2020backdoor,xi2020graph} in various domains such as image, text, and graph. For instance, Gu et al.~\cite{gu2017badnets} proposed BadNets where an attacker can inject a backdoor into a classifier via poisoning its training dataset. In particular, the goal of the attacker is to make the backdoored classifier predict an attacker-chosen target label for any testing inputs embedded with an attacker-chosen trigger. To reach the goal, the attacker adds a backdoor trigger to a certain fraction of training inputs and relabels them as the target label. Dai et al.~\cite{dai2019backdoor} proposed backdoor attacks against LSTM-based text classification via poisoning its training dataset. In particular, they showed that the backdoored classifier classifies any text embedded with a pre-defined, attacker-chosen trigger into the attacker-chosen target class. Zhang et al.~\cite{zhang2020backdoor} developed backdoor attacks to graph neural networks. In particular, a backdoored GNN classifier predicts a target label for a testing graph embedded with an attacker-chosen subgraph. Backdoor attacks have also been studied for pre-trained encoders in self-supervised learning~\cite{jia2021badencoder,carlini2021poisoning,zhang2021trojaning}. Next, we will respectively introduce those studies. 

Recently, two studies~\cite{jia2021badencoder,carlini2021poisoning} developed backdoor attacks to pre-trained image encoders. 
Jia et al.~\cite{jia2021badencoder} proposed BadEncoder. In particular, BadEncoder crafts a backdoored image encoder with multiple backdoors from a clean one via solving an optimization problem. When the backdoored image encoder is used to build downstream classifiers for downstream tasks, each downstream classifier will predict a target label for any inputs embedded with a backdoor trigger. Carlini et al.~\cite{carlini2021poisoning} developed backdoor attacks to image encoders pre-trained on (image, text) pairs. In particular, an attacker aims to make any testing inputs embedded with an attacker-chosen backdoor trigger be classified as an attacker-chosen target label via poisoning the pre-training dataset. Backdoor attacks have also been studied for pre-trained language models~\cite{zhang2021trojaning}. For instance, Zhang et al.~\cite{zhang2021trojaning} developed trojan attacks to language models such as BERT~\cite{devlin2018bert} and GPT-2~\cite{radford2019language}. In particular, they showed that an attacker can inject a backdoor into a language model such that its outputs for testing inputs embedded with a backdoor trigger are attacker desired. To the best of our knowledge, backdoor attacks to graph encoders haven't been explored yet. We believe it is an interesting future work to study backdoor attacks to graph encoders. The key challenge is how to exploit the unique characteristics of graphs.

Many empirical defenses~\cite{liu2018fine,wang2019neural,xu2019detecting} and certified defenses~\cite{chiang2019certified,wang2020certifying,zhang2020backdoor,levine2020randomized} have been developed to mitigate backdoor attacks to classifiers. However, those defenses are insufficient to defend against backdoor attacks to pre-trained encoders. For instance, Jia et al.~\cite{jia2021badencoder} found that existing defenses against backdoor attacks to classifiers are insufficient to mitigate BadEncoder. 
Generally speaking, there are three defense strategies: prevention, detection, and recovery. To prevent backdoor attacks, one can use encoders from trusted sources or try to obtain a clean pre-training dataset (e.g., filtering poisoned training examples) to pre-train an encoder. Given a pre-trained encoder, we can also try to detect whether it is backdoored or not. The detection of backdoored encoder is essentially a binary classification problem. Therefore, the key challenge is how to achieve a good tradeoff between false positive rate and false negative rate. Once a pre-trained encoder is predicted to be backdoored, we can try to develop recovery methods to erase the backdoor in it. 

\myparatight{Poisoning Attacks} Poisoning attacks aim to corrupt a machine learning system via poisoning its training dataset (known as \emph{data poisoning attacks})\cite{nelson2008exploiting,rubinstein2009antidote,biggio2012poisoning} or its training process (known as \emph{model poisoning attacks})~\cite{bhagoji2019analyzing,fang2020local}. The corrupted machine learning system may produce inaccurate outputs for indiscriminate testing inputs (a.k.a. \emph{untargeted poisoning attacks}); and the corrupted  system may produce attacker-chosen outputs for attacker-chosen testing inputs while its outputs for other testing inputs are unaffected (a.k.a. \emph{targeted poisoning attacks}). 
Data poisoning attacks have been extensively studied for image classification models~\cite{biggio2012poisoning,munoz2017towards,shafahi2018poison,suciu2018does,demontis2019adversarial,zhu2019transferable,huang2020metapoison}, recommender systems~\cite{li2016data,yang2017fake,fang2018poisoning,fang2020influence,huang2021data},  graph-based classification~\cite{zugner2018adversarial,wang2019attacking,jia2020certifiedcom}, and so on~\cite{nelson2008exploiting,rubinstein2009antidote}. Model poisoning attacks have been studied for federated learning systems~\cite{bhagoji2019analyzing,fang2020local,bagdasaryan2020backdoor} and local differential privacy protocols~\cite{cao2021data,cheu2021manipulation}.  However,  poisoning attacks, both untargeted and targeted, are largely unexplored for self-supervised learning. For instance, it is still an open question about how to poison the pre-training dataset such that the downstream classifiers built upon the poisoned encoders have low classification accuracies. We note that a recent work~\cite{carlini2021poisoning} proposed  targeted data poisoning attack to contrastive learning. Roughly speaking, an attacker aims to poison the pre-training dataset such that downstream classifiers predict an attacker-chosen target label for an attacker-chosen testing input. However, their attack is only applicable to the scenario where the pre-training dataset contains (image, text) pairs. 

To mitigate poisoning attacks, many empirical defenses~\cite{rubinstein2009antidote,barreno2010security,biggio2011bagging,steinhardt2017certified,tran2018spectral,jagielski2018manipulating} and certified defenses~\cite{ma2019data,rosenfeld2020certified,jia2020intrinsic,levine2020deep,jia2020certified} have been proposed. It is unclear whether those defenses can be generalized or are effective when generalized to defend against poisoning attacks to self-supervised learning,  which is an interesting future direction. Another future direction is to develop new defenses along three general defense strategies (i.e., prevention, detection, and recovery) against  poisoning attacks to self-supervised learning. For instance, it would be an interesting future work to detect and recover poisoned encoders.

\myparatight{Evasion Attacks} In evasion attacks, an attacker aims to add adversarial perturbations to testing inputs such that the outputs of a machine learning system for the adversarially perturbed inputs are desired by the attacker. For instance, many studies~\cite{Szegedy14,goodfellow2014explaining,madry2017towards,carlini2017towards,zhang2019theoretically} showed that an attacker can craft a human-imperceptible, small perturbation for an input such that a machine learning classifier predicts an attacker-chosen label or a wrong label when the perturbation is added to the input. Both empirical defenses~\cite{goodfellow2014explaining,madry2017towards,cao2017mitigating} and certified defenses~\cite{wong2017provable,raghunathan2018certified,lecuyer2018certified,cohen2019certified,jia2019certified,jia2020almost} were developed to defend against evasion attacks. Among the empirical defenses, adversarial training~\cite{goodfellow2014explaining,madry2017towards} was shown to be the most effective one against evasion attacks. Roughly speaking, adversarial training crafts adversarially perturbed inputs for training inputs and uses  them to train a classifier in each training epoch. Randomized smoothing~\cite{lecuyer2018certified,cohen2019certified} is a state-of-the-art certified defense which is scalable to large-scale neural networks and applicable to arbitrary classifiers. In particular, given a classifier and a testing input, randomized smoothing randomizes the input (e.g., add random Gaussian noise to the input)  and estimates the probability that the given classifier predicts each label for the randomized input. The label with the largest probability is predicted for the testing input by randomized smoothing. The predicted label is provably unchanged when the adversarial perturbation added to the input is bounded.  

Some recent studies~\cite{chen2020adversarial,kim2020adversarial,jiang2020robust} proposed to pre-train robust encoders by generalizing adversarial training~\cite{madry2017towards}. Roughly speaking, the idea is to craft adversarially perturbed inputs that incur large self-supervised learning loss and then use them to pre-train an encoder. Similar to adversarial training for classifiers~\cite{madry2017towards}, this essentially can be formulated as a minimax optimization problem. Existing studies~\cite{chen2020adversarial,kim2020adversarial} found that downstream classifiers built upon encoders pre-trained using adversarial training are much more robust than those built upon encoders that do not use adversarial training. An interesting future work along this direction is to design new adversarial training methods to train empirically robust encoders such that downstream classifiers built upon them are more robust. Another future direction is to develop provably robust encoders such that downstream classifiers built upon them are provably robust. For instance, generalizing existing techniques developed for  classifiers to build provably robust encoders would be an interesting future work.

\section{One Problem on Availability}
\label{availability}
Untargeted poisoning attacks, which damage the utility of a pre-trained encoder, can be viewed as compromising the availability of foundation models since an ineffective encoder won't be used and is eventually unavailable to users, leading to denial-of-service. Next, we discuss another availability attack called \emph{Resource Depletion Attack} to foundation models.

\myparatight{Resource Depletion Attacks} In resource depletion attacks, an attacker aims to poison the training dataset or perturb a testing input such that more resources (e.g., computation resources) are consumed when the poisoned training dataset is used to train a machine learning model or the perturbed testing input is fed into a machine learning model. Resource depletion attacks have been studied for classifiers~\cite{shumailov2020sponge,hong2020panda}. For instance, Hong et al.~\cite{hong2020panda} showed that an attacker can craft some perturbed inputs to increase inference time or consumed energy of a machine learning classifier deployed on resource-constrained devices such as IoT devices. 

An attacker can also perform resource depletion attacks to foundation models either in training or testing phase. In particular, self-supervised learning usually leverages a large amount of unlabeled data collected from the Internet to pre-train an encoder. Therefore, an attacker can poison the training dataset by uploading poisoned data to the Internet to reduce the convergence rate and increase the time of pre-training an encoder. 
An attacker could also carefully perturb inputs to increase the inference time of a deployed pre-trained encoder. 
\section{Conclusion}
\label{conclusion}
Foundation model is commonly believed to be a promising approach for general-purpose AI. In this book chapter, we discuss 10 key security and privacy problems of foundation models from the perspectives of confidentiality, integrity, and availability. We hope our work can inspire future research on uncovering the security and privacy risks of foundation models as well as enhancing their security and privacy in adversarial settings.

%
\section*{Acknowledgements} 
This work is supported by the National
Science Foundation under Grants No. 1937786 and
the Army Research Office under Grant No. W911NF2110182. Any
opinions, findings and conclusions or recommendations expressed
in this material are those of the author(s) and do not necessarily
reflect the views of the funding agencies.

{
\balance{
\bibliographystyle{unsrt}
\bibliography{refs}
}}

\end{document}